\documentclass[%
preprint,
 longbibliography,
 amsmath,amssymb,
 aps,
 pra,
]{revtex4-1}

\usepackage{graphicx}% Include figure files
\usepackage{dcolumn}% Align table columns on decimal point
\usepackage{bm}% bold math
\usepackage{mathtools}
\usepackage{comment}
\usepackage{afterpage}
\usepackage{color}

\begin{document}

\preprint{APS/123-QED}

\title{Self-assembly of active amphiphilic Janus particles}
\author{S. A. Mallory$^1$,  F.Alarcon$^{2}$, A. Cacciuto$^{1*}$, C. Valeriani$^{2}$}
\email{ac2822@columbia.edu}
\email{cvaleriani@ucm.es}
\affiliation{$^{1}$Department of Chemistry, Columbia University\\ 3000 Broadway, New York, NY 10027\\ }
\affiliation{$^{2}$Departamento de Fisica Aplicada I, Facultad de Ciencias Fisica, Universidad Complutense de Madrid, 28040 Madrid, Spain}

\date{\today}% It is always \today, today,

\begin{abstract}
 
In this article, we study the phenomenology of a two dimensional dilute suspension of active amphiphilic Janus particles.  We analyze how the morphology of the aggregates emerging from their self-assembly depends on the strength and the direction of the active forces. We systematically explore  and contrast the phenomenologies resulting from particles with a range of attractive patch coverages. Finally, we illustrate how the geometry of the colloids and the directionality of their interactions can be used to control the physical properties of the assembled active aggregates and suggest possible strategies to exploit self-propulsion as a tunable driving force for self-assembly.  

\end{abstract}

\maketitle

%\tableofcontents

\section{\label{sec:intro}Introduction}

Active matter continues to be one of the most exciting new fields in statistical mechanics and materials engineering. Whether one focuses on the chaotic dynamics of human crowds or the peculiar swarming of swimming bacteria, a common feature of all active systems is their ability to convert environmental or internal energy into systematic movement.  The inherent self-propulsion or activity at the ``single particle level" in these systems gives rise to a remarkable range of different collective behaviors. For a number of examples and a thorough discussion on the topic, we direct the reader to a number of comprehensive reviews (see for instance \cite{bechinger_active_2016,wang_one_2015,toner_hydrodynamics_2005,cates_motility-induced_2015,ramaswamy_mechanics_2010,klapp_collective_2016,marchetti_minimal_2016,bialke_active_2015,zottl_emergent_2016,dey_chemically_2017,wang_small_2013,menzel_tuned_2015,marchetti_hydrodynamics_2013,j.ebbens_pursuit_2010,chate_modeling_2008,romanczuk_active_2012,j.ebbens_pursuit_2010} and the references therein). In this work, we are predominately concerned with colloidal active systems, and developing new methods to use activity as a tool to optimize colloidal self-assembly. Synthetic microswimmers are expected to fill an important void within the material science community as they possess the rather unique ability of being able to manipulate, sense, and transport material at the micro-scale.  The potential applications are far reaching and include: targeted drug delivery \cite{din_synchronized_2016}, information storage and computation \cite{woodhouse_active_2017}, and the clean-up and neutralization of environmental pollutants \cite{ebbens_active_2016}. These applications are all built around the unique self-driven nature of the microswimmers and in many ways, active colloids look to serve as the perfect building blocks for the self-assembly of the next generation of functional, environment-sensing microstructures able to perform specific tasks in an autonomous and targeted manner. 

Through the groundbreaking work of synthetic chemists and material scientists, there is now a continually growing library of synthetic microswimmers and active colloids. These active particles are the synthetic analogs of swimming bacteria.  However, a major benefit of these synthetic variants over swimming bacteria is that both the inter-particle interactions and the swimming velocity can be systematically controlled. For details about the growing variety of synthetic microswimmers and their associated propulsion mechanisms, we direct the readers to several recent reviews \cite{wang_one_2015,wang_small_2013,bechinger_active_2016,j.ebbens_pursuit_2010}. The majority of the published work on colloidal active matter focuses on suspensions of active particles which interact via an isotropic potential (whether it be attractive \cite{mognetti_living_2013,redner_reentrant_2013,prymidis_self-assembly_2015,schwarz-linek_phase_2012}, or purely repulsive \cite{stenhammar_activity-induced_2015,stenhammar_phase_2014,tailleur_statistical_2008,theurkauff_dynamic_2012,speck_collective_2016,speck_effective_2014,buttinoni_dynamical_2013}).  Only recently, has the field branched out to explore the interplay between activity and anisotropic pairwise interactions. We are only now developing a systematic understanding of how active forces can be exploited in conjunction with anisotropic pair interactions to design macroscopic assemblies with desired structural properties (see \cite{mallory_activity-assisted_2016,di_leonardo_active_2016,zhang_directed_2016,singh_non-equilibrium_2017,johnson_dynamic_2017,gao_dynamic_2017,pu_reentrant_2017} and references therein). 

The inspiration and motivation for this work was derived from the recent experimental synthesis of active colloids with directed anisotropic interactions. By altering the surface chemistries of an active colloid, it is possible to introduce a number of different directed anisotropic  interactions. In a recent publication, Yan et al. \cite{yan_reconfiguring_2016}  presented a general strategy, that exploits electrostatic imbalance between the two hemispheres of an active dipolar Janus particle, to reconfigure their assembly into a number of different collective states. Several research groups \cite{kaiser_active_2015,guzman-lastra_fission_2016} have also begun to probe the clustering behavior of a small number of dipolar Janus particles with a specific focus on analyzing how the structural properties of the self-assembled clusters are affected by the introduction of self-propelling forces. An alternative approach to introducing directed anisotropic  interactions is to pattern the hydrophobic region on the surface of the active colloid \cite{gao_organized_2013}. Unlike dipolar active Janus particles that consists of two hemispheres of opposite charge, amphiphilic active patchy particles (APPs) consist of a polar and a hydrophobic region. A recent experimental realization of an active amphiphilic patchy particle consists of silica microspheres passivated with a hydrophobic ligand (octadecyltrichlorosilane, OTS), whose hemisphere is then covered by a Pt cap. These active APPs experience a short ranged attraction between the hydrophobic domains, while the interaction between the Pt capped hemispheres is purely repulsive. In essence, the Pt cap restricts the inter-particle attraction to the hydrophobic region which gives rise to well defined assemblies.  By tuning the size of the hydrophobic domain and that of the Pt cap, it is possible to control the angular range of the interaction, and thus the number of colloids that can simultaneously stick to one another.

In this paper, we report the self-assembly dynamics of a two dimensional dilute suspension of active APPs and study the structural and morphological properties of the aggregates that spontaneously form in the presence of active forces. We consider active APPs with different sizes of the hydrophobic regions spanning from 25\%, 50\% and 75\% of the total area of the particle, and for two different propelling directions: one that is in the same direction as the hydrophobic patch and one that is in the opposite direction of the hydrophobic patch. Our results unveil a nontrivial morphological dependence of the self-assembled structures on the active forces and hydrophobic coverage.

\section{\label{sec:model}Model}

\begin{figure}[h!]
 \centering
    \includegraphics[width=0.8\textwidth]{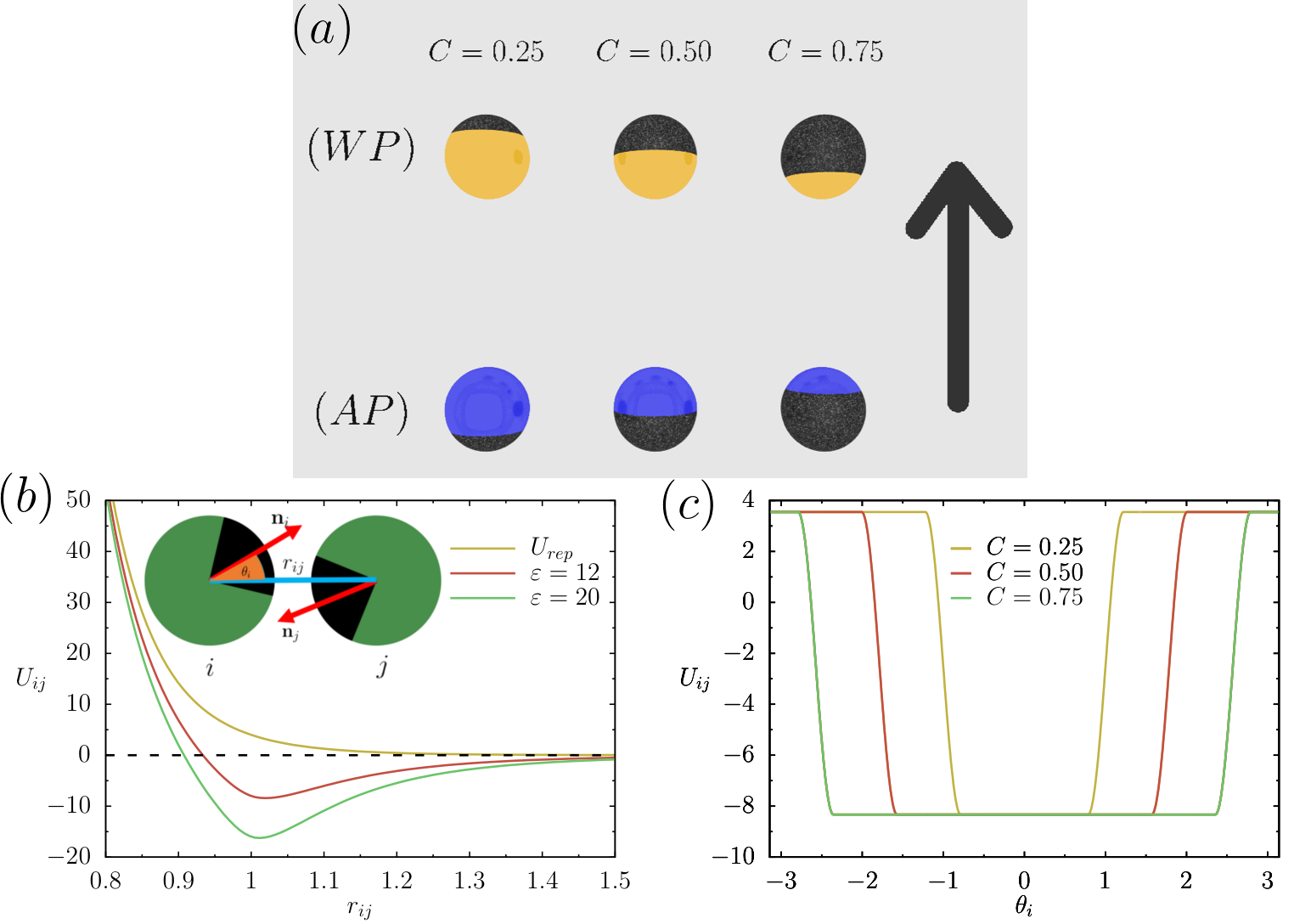}
  \caption{(a) Illustration of the three patch coverages considered in this study. The hydrophobic patch is shown in black with percentage coverage $C$ corresponding to, from left to right, $C=0.25$, $C=0.50$, and $C=0.75$, respectively.  The large arrow indicates the propulsion direction.  The top row corresponds to patchy colloids propelled in the direction of the patch $(WP)$, whereas the bottom row  against the patch $(AP)$.  (b) Potential $U_{ij}$ given as a function of the separation distance $r_{ij}$. The potential for the two primary binding energies $\varepsilon=12$ and $\varepsilon=20$ are shown, as well as the purely repulsive component of the potential. (c) Potential $U_{ij}$ given as a function of the angle $\theta_i$ for the various patch coverages considered in this study. }
  \label{fig:1}
\end{figure}

In this manuscript, we consider three different attractive patch coverages, $C$, defined as the ratio between the area of the hydrophobic patch and the total area of the particle. In Figure \ref{fig:1}, we illustrate from left to right the different attractive patch coverages: $C=0.25$, $C=0.50$, and $C=0.75$. The large arrow in the figure indicates the direction of self-propulsion with respect to the attractive patch. The top row corresponds to APPs that are propelled in the same direction as the patch, $\pmb{\eta_i}=\pmb{n_i}$, or with the patch, which we abbreviate as $(WP)$.  The bottom row corresponds to APPs propelled against the attractive patch, $\pmb{\eta_i}=-\pmb{n_i}$, which is abbreviated as $(AP)$. In Fig~\ref{fig:1}, the black portion of the particles correspond to the attractive patches, and we use yellow and blue to differentiate between the $(WP)$ and $(AP)$ cases, respectively. For each patch coverage, we will first study the behavior of the passive counterpart at the same density, represented as green (repulsive patch) / black (attractive patch) colloids.

We simulate a two dimensional dilute suspension of $N=2000$  amphiphilic patchy particles (APPs) at  an area fraction of $\phi = 0.1$. Each APP is modeled as a disk with diameter $\sigma$, and undergoes Brownian dynamics at a constant temperature $T$ according to the coupled translational and rotational equations:

\begin{equation}
\label{tra}
\dot{\pmb{r}}(t) = \frac{1}{\gamma} \pmb{F}(\{\pmb{r}_{ij},\pmb{\eta}_i,\pmb{\eta}_j\}) +    \upsilon_p \, \pmb{\eta}(t) + \sqrt{2D}\,\pmb{\xi}(t)
\end{equation}

\begin{equation}
\label{rot}
\dot{\pmb{\eta}}(t) = \frac{1}{\gamma_r}\pmb{T}(\{\pmb{r}_{ij},\pmb{\eta}_i,\pmb{\eta}_j\}) + \sqrt{2D_r}\, \pmb{\xi}(t) \times \pmb{\eta}(t)
\end{equation}

\noindent Activity or self-propulsion is introduced through a directional propelling velocity of constant magnitude $\upsilon_p$ and is directed along a predefined orientation vector $\pmb{\eta}$.  $\pmb{\eta}$ is centered at the origin of each particle and is restricted to rotate in the 2D plane of the system.
%There are no torques, thermal or conservative, that would cause the orientation vector to rotate out of the plane of the system.} 
The translational diffusion coefficient $D$ is related to the temperature $T$ and the translational friction $\gamma$ via the Stokes-Einstein relation $D=k_{\rm B}T/\gamma$. We make the typical assumption that the rotational diffusion coefficient $D_r=k_{\rm B}T/\gamma_r$ satisfies the relation $D_r = (3D)/\sigma^2$. Thus, the P\'eclet number can be defined as $Pe=\sigma v_p/D$.
The  solvent induced Gaussian white-noise terms for both the translational and rotational motion are characterized by $\langle \pmb{\xi}(t)\rangle = 0$ and $\langle \xi_i(t) \xi_j(t^\prime)\rangle = \delta_{ij}\delta(t-t^\prime)$. The conservative inter-particle forces and torques acting on each colloid are given by $\pmb{F}(\{\pmb{r}_{ij},\pmb{\eta}_i,\pmb{\eta}_j\})$ and  $\pmb{T}(\{\pmb{r}_{ij},\pmb{\eta}_i,\pmb{\eta}_j\})$ respectively, and can be computed as 

\begin{equation}
\pmb{F}(\{\pmb{r}_{ij},\pmb{\eta}_i,\pmb{\eta}_j\}) = \sum_j \pmb{F}_{ij} = -\sum_j \pmb{\nabla}_{\pmb{r}_{ij}} U_{ij}\left(r_{ij}, \theta_i,\theta_j \right)
\end{equation}
and 
\begin{equation}
%\pmb{T}(\{\pmb{r}_{ij},\pmb{\eta}_i,\pmb{\eta}_j\}) = \sum_j \pmb{T}_{ij} = -\sum_j \pmb{\nabla}_{\pmb{n}_j} U_{ij}
\pmb{T}(\{\pmb{r}_{ij},\pmb{\eta}_i,\pmb{\eta}_j\}) = \sum_j \pmb{T}_{ij} = \sum_j (\pmb{\eta}_i \times  \dfrac{\partial U_{ij}\left(r_{ij}, \theta_i,\theta_j \right)}{\partial \pmb{\eta}_i})
\end{equation}

\noindent where $U_{ij}$ is the interaction potential between colloids. The anisotropic pair potential is given by

\begin{equation}
U_{ij}\left(r_{ij}, \theta_i,\theta_j \right) = U_{rep}(r_{ij}) 
+ U_{att}(r_{ij}) \phi(\theta_i) \phi(\theta_j )
\end{equation}

\noindent where
\begin{equation}
U_{rep}(r_{ij}) = 
\begin{dcases} 
      4 \varepsilon_{rep} \left(\frac{\sigma}{r_{ij}}\right)^{12} & r_{ij}\leq 1.5 \sigma \\
      0 & r_{ij} >  1.5 \sigma  
\end{dcases}
\end{equation}

\noindent and

\begin{equation}
U_{att}(r_{ij}) = 
\begin{dcases} 
      4 \varepsilon \left[ \left(\frac{\sigma/2}{r_s + \sigma /2 \times  2^{1/6}} \right)^{12} - \left(\frac{\sigma/2}{r_s + \sigma /2 \times  2^{1/6}}\right)^6 \right] & r_{ij}\leq 1.5\sigma \\
      0 & r_{ij} > 1.5\sigma  
\end{dcases}
\end{equation}

\noindent Here $\varepsilon$ is the binding energy, $r_{ij}$ is the distance between the centers of particles $i$ and $j$, while $r_s=|r_{ij}-\sigma|$ is the distance between the particles surfaces. Here, we choose a value of $\varepsilon_{rep}=1$ for the repulsive component of the potential. The potential $U_{ij}$ as a function of the separation distance $r_{ij}$ is given in Figure \ref{fig:1}(b) for different pair configurations.

The angular interaction $\phi(\theta_i)$ is given by 

\begin{equation}
\label{angles}
 \phi (\theta_i) = 
\begin{dcases} 
      1 & \theta_i \leq \theta_{max} \\
      \cos^2\left(\frac{\pi (\theta_i - \theta_{max} )}{2 \theta_{tail}} \right) & \theta_{max} \leq \theta_i \leq \theta_{max} + \theta_{tail} \\
      0 & \mbox{otherwise}
\end{dcases}
\end{equation}

\noindent where $\theta_i$  is defined as the angle between the patch unit vector $\pmb{n}_i$ and the inter-particle vector $\pmb{r}_{ji}=\pmb{r}_j-\pmb{r}_i$, hence $\cos(\theta_i)=\pmb{r}_{ji} \cdot \pmb{n}_i$ (similarly $\theta_j$ is   the angle between patch vector $\pmb{n}_j$ and inter-particle vector $\pmb{r}_{ij} = - \pmb{r}_{ji}$). The corresponding expression for $\phi(\theta_j)$ is also given by Eq. 8. The angular potential $\phi$ is a smooth step function that modulates the angular dependence of the potential and is equal to 1 within the region $\theta_i < \theta_{max}$ and decays to zero following the expression above. The percentage coverage $C=\{ 0.25,0.5,0.75\}$ is related with $\theta_{max}=\{45,90,135\}$ degrees, respectively.
The particular value of $\theta_{tail} =25$ degrees has been selected to generate a sufficiently smooth potential at the Janus interface. The potential $U_{ij}$ is shown in Fig.~\ref{fig:1}c as a function of the angle $\theta_i$ for two particles at contact while the orientation of particle $j$ is held fixed.  This potential was adapted from the model introduced by Miller \& Cacciuto \cite{miller_hierarchical_2009}, and is appropriate to describe the behavior of patchy particles at high salt concentration.
In fact, the nature of the interaction between two Janus particles is very much dependent on the degree of electrostatic screening of the polar hemisphere of the particle. As observed in experiments\cite{hong_clusters_2008},
when unscreened, the polar side of the particles would provide a significant angular dependence on the orientation of the particles, and even prevent their aggregation\cite{hong_clusters_2008}.
%In this limit, the charge on the hydrophilic side of the colloid is well screened and {\color{red} can be mostly ignored}\cite{hong_clusters_2008}.
In our simulations $\sigma$ and $k_{\rm B}T$ are used as the basic units of length and energy, while $\tau=\sigma^2/D$ is our unit of time. All simulations were run for a minimum of $10^7$ steps with time step $\Delta t=10^{-5}\tau$.
\section{\label{results}Results}

\subsection{Dilute suspension of active APPs with patch coverage $(C=0.25)$}

\begin{figure}[h!]
 \centering
    \includegraphics[width=0.8\textwidth]{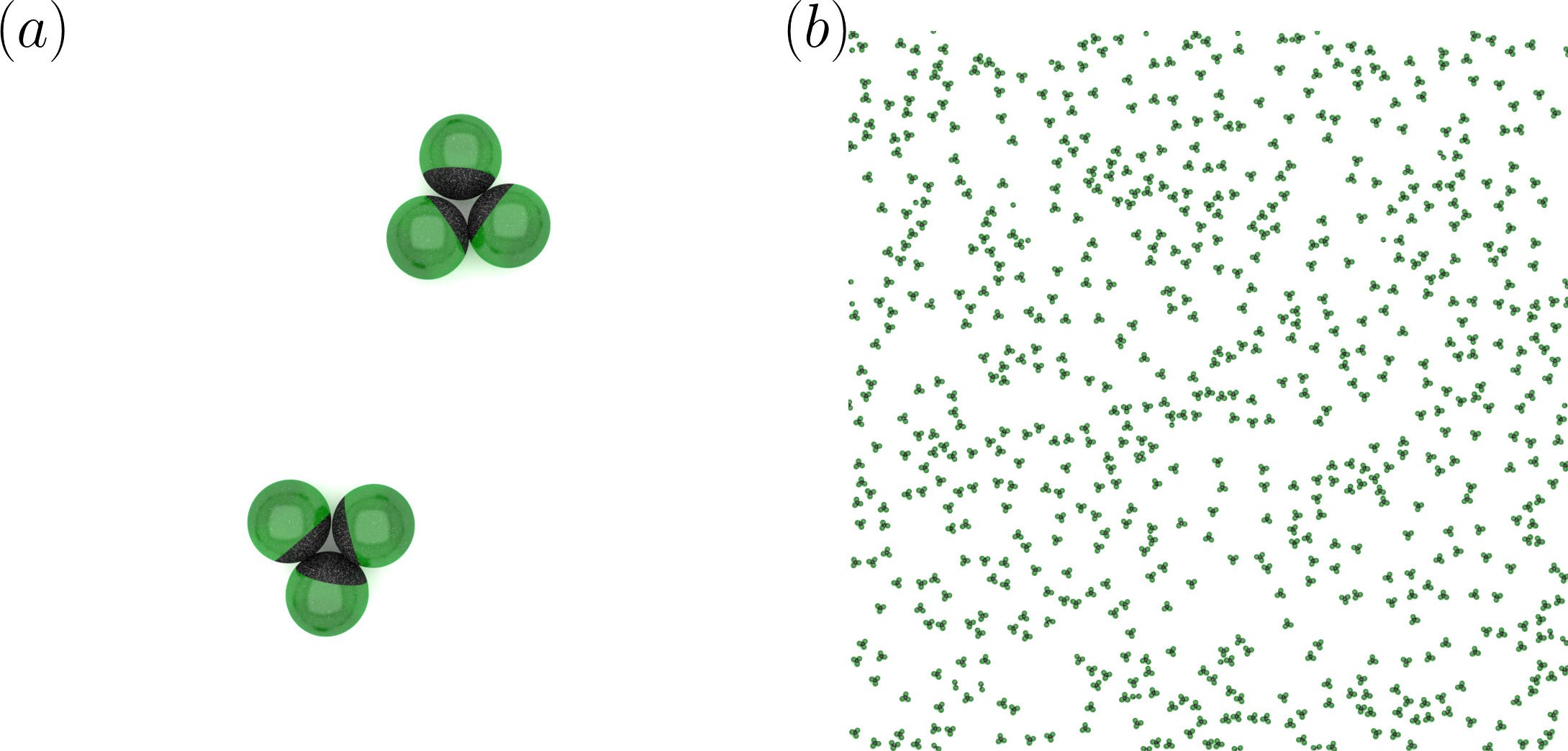}
  \caption{(a) Snapshot of the stable trimer structure for passive suspension of APPs with patch coverage $C=0.25$. (b) Typical configuration observed for binding energy $\varepsilon=20$.}
  \label{fig:2}
\end{figure}

We begin by considering the bulk properties of a low density suspension of passive APPs with a small attractive patch. For all patch coverages, the passive system serves as a reference to the equilibrium behavior. In the passive system at $C=0.25$, the maximum number of attractive interactions or bonds a given APP can form is limited to two, resulting in clusters of three APPs at most. In this configuration the APPs arrange into equilateral triangles with the attractive patch oriented toward the barycenter of the triangle as shown in Fig.~\ref{fig:2}(a). A snapshot of a typical configuration at large binding energy ($\varepsilon=20$) is given in Figure \ref{fig:2}(b). As one would expect, the binding energy determines the clustering behavior of these suspensions, and the shape of the largest possible stable cluster is determined by the geometric constraints imposed by the shape of the attractive patch: a trimer in this small patch case ($C=0.25$).

\begin{figure}[h!]
 \centering
    \includegraphics[width=0.8\textwidth]{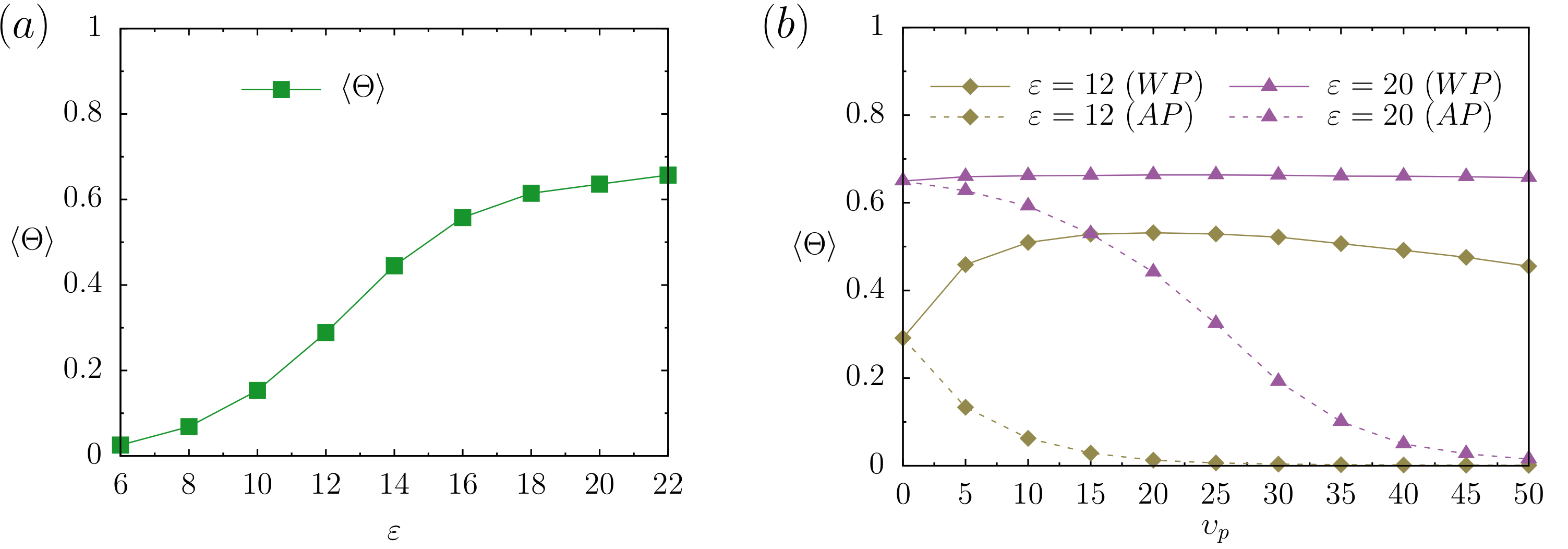}
  \caption{(a) Average degree of aggregation $\langle \Theta \rangle$ as a function of $\varepsilon$ for passive suspensions with $C=0.25$. (b) Average degree of aggregation $\langle \Theta \rangle$ as a function of $\upsilon_p$ for active suspensions with $C=0.25$. Two different biniding energies are considered: $\varepsilon=12$ (light green) and $\varepsilon=20$ (magenta). Solid lines correspond to the (WP) case, while the (AP) case is indicated by the dotted lines.}
  \label{fig:3}
\end{figure}

To study how these particles self-assemble, we compute the average degree of aggregation $\langle \Theta \rangle$ of the system defined as in Ref.\cite{mognetti_living_2013,prymidis_self-assembly_2015}: 

\begin{equation}
\langle \Theta \rangle = 1 - \frac{ \langle N_{c} \rangle } { N } = 1 - \frac{1}{\langle S \rangle }
\end{equation}

\noindent where $\langle N_{c} \rangle$ is the average number of clusters in the system and $ \langle S \rangle$ is the average number of APPs in a cluster.  For a system, that has aggregated into a single cluster $\langle \Theta \rangle = 1$, and conversely a system composed of single particle clusters will give a value $\langle \Theta \rangle = 0$. In Figure \ref{fig:3}(a), we show how the average degree of aggregation $\langle \Theta \rangle$  depends on the binding energy $\varepsilon$. For small binding energies the system mainly consists of monomers and dimers, while for large $\varepsilon$ it almost exclusively consists of trimers. For the largest binding energy, we find that $\langle \Theta \rangle \simeq 0.66$, which is consistent with the theoretical maximum value of $\langle \Theta \rangle = 2/3$ occurring when all aggregates are trimers.
 
With an understanding of the equilibrium behavior, we now turn our attention to the effect of self-propulsion. Specifically, we consider the two cases where APPs are either propelled in the direction of the attractive patch $(WP)$ or in the opposite direction $(AP)$. For these active suspensions, we only consider two different values of the binding energy, $\varepsilon=12$ and $\varepsilon=20$, and a range of activities up to $\upsilon_p=50$. Fig.~\ref{fig:3}(b) summarizes the results of this study for $C=0.25$. Interestingly, the clustering behavior of active APPs is highly sensitive to the direction of self-propulsion relative to the attractive patch. At the smaller binding energy $\varepsilon=12$, thermal fluctuations in the passive case are able to easily break the bond between two APPs. As a result, $\langle \Theta \rangle$ in the passive system is quite small and takes a value of $\langle \Theta \rangle=0.3$. As we introduce self-propulsion in the $(WP)$ direction, we observe  an increase in  $\langle \Theta \rangle$ to values above $\langle \Theta \rangle=0.50$. Here activity stabilizes the formation of dimers and trimers structures. Whereas the opposite occurs when the propulsion is in the $(AP)$ direction, where we observe $\langle \Theta \rangle$ rapidly dropping to nearly zero where even small clusters quickly break into unpaired monomers. This behavior is even more pronounced at higher binding energies. At $\varepsilon=20$, the system mainly consists of trimers for the $(WP)$ direction. Interestingly, since in  this configuration, the 
net sum of the active forces is zero, these trimers  are essentially inert,  resulting in a low density fluid where each trimer undergoes Brownian diffusion. Once more, in the $(AP)$ direction we observe $\langle \Theta \rangle$  dropping to nearly zero (even though slower than in the $\varepsilon=12$ case).

This behavior is particularly interesting as it further reinforces the notion that activity in the $(WP)$ direction can be used to strengthen particle-particle interactions and significantly improves self-assembly of compact structures~\cite{mallory_improving_2017}. It is worth pointing out, however, that the small lowering in $\langle \Theta \rangle$ for the largest activities at $\varepsilon=12$ is due to either collisions of a trimer with fast moving unpaired APPs, or the formation of unstable configurations where the vectorial sum of the propelling axes in the trimer deviates sufficiently from zero to generate internal torques able to break the trimer. These destructive events are very rare for $\varepsilon=20$ as most APPs occupy a stable configuration within a trimer, and thus we observe no significant decline with increasing $\upsilon_p$ (at least within the range of velocities considered in this study). These trimer destroying events occur more frequently at lower binding energies, as there are more unpaired APPs and the interaction energy is not as effective at maintaining fairly strict angular orientations within the trimer. 

Although, it is tempting to think of the role of activity in this system as having the simple effect of shifting the strength of the inter-particle interactions as $\varepsilon_{\rm eff}=\varepsilon\pm \alpha \upsilon_p$,
with the plus or minus corresponding to the $(WP)$ and $(AP)$ directions respectively, and where $\alpha$ is a phenomenological constant, the reality is  more complicated. In fact, while on the one hand, self-propulsion can either strengthen or weakens the interaction energy of the particles in a cluster,  on the other hand self-propulsion increases the translational velocity of the unpaired particles in the bulk (that one could think of as an effective "temperature"), which makes a straightforward mapping to an effective equilibrium system rather nontrivial.

\subsection{Dilute suspension of active APPs with patch coverage $(C=0.50)$}

\begin{figure}[h!]
 \centering
    \includegraphics[width=0.8\textwidth]{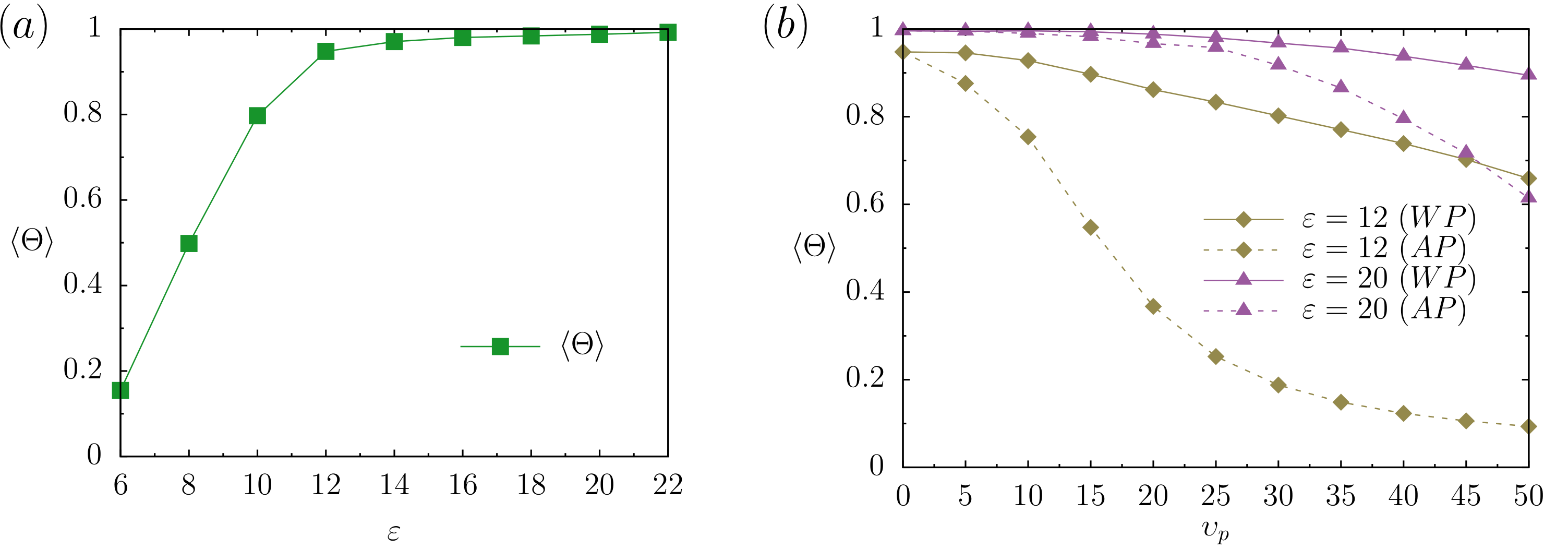}
  \caption{(a) Average degree of aggregation $\langle \Theta \rangle$ as a function of $\varepsilon$ for passive suspensions with $C=0.50$. (b) Average degree of aggregation $\langle \Theta \rangle$ as a function of $\upsilon_p$ for active suspensions with $C=0.50$. Two different binding energies are considered: $\varepsilon=12$ (light green) and $\varepsilon=20$ (magenta). Solid lines correspond to the (WP) case, while the (AP) case is indicated by the dotted lines.}
  \label{fig:4}
\end{figure}

We now consider the behavior of APPs with patch coverage $C=0.50$. For this  patch coverage, the maximum number of bonding interactions for a given APP is limited to four. For this reason, unlike the $C=0.25$ case, there isn't a simple compact structure determining the maximum cluster size for a given set of system parameters (i.e. density, binding energy, temperature, etc). In the passive system, the most stable local configuration is achieved when each APP is located at the vertices of a parallelogram with the attractive patches oriented inward toward one another. This results in the most abundant macrostructures being two-particle thick, rigid chain-like clusters, whose length is controlled by the binding energy $\varepsilon$ (see top panels in Figures \ref{fig:5} and \ref{fig:6}). For small values of $\varepsilon$ (top panel of Fig.~\ref{fig:5}), clustering is limited to the dynamic assembly and disassembly of chain like clusters. In this regime, predominately linear chain structures are observed and the typical number of APPs in a single cluster is on the order of 20. As expected, the length of these chains increases with increasing binding energy. However, as the binding energy increases to large values, branches and kinks develop along the chain backbone (top panel of Fig.~\ref{fig:6}). For the largest binding energies considered, we observe the formation of large branched networks spanning the size of the system. This qualitative description of the behavior of the passive system can be  represented quantitatively with average degree of aggregation $\langle \Theta \rangle$ shown in Fig~\ref{fig:4}(a).

\afterpage{%
\begin{figure}[h!]
 \centering
    \includegraphics[width=0.8\textwidth]{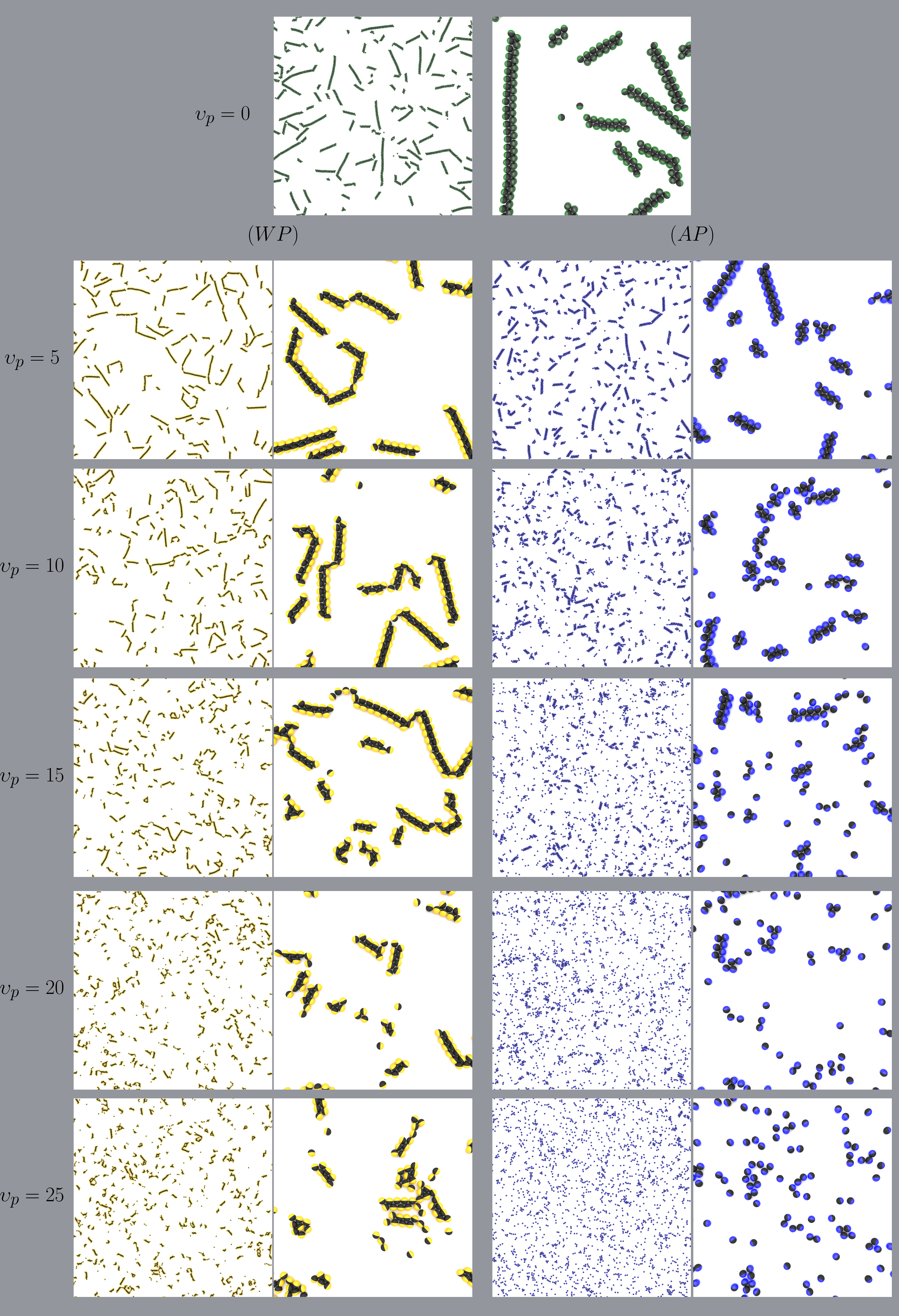}
  \caption{Typical simulation snapshot for active patchy particles with patch coverage $C=0.50$ and $\varepsilon=12$. The passive reference system $\upsilon_p=0$ is shown at the top of the figure.  The (WP) case is shown in the left column while the (AP) case is shown in the right column.  Each system snapshot is accompanied by an enlarged snapshot highlighting the local cluster geometry. }
  \label{fig:5}
\end{figure}
\clearpage
}
\afterpage{%
\begin{figure}[h!]
 \centering
    \includegraphics[width=0.8\textwidth]{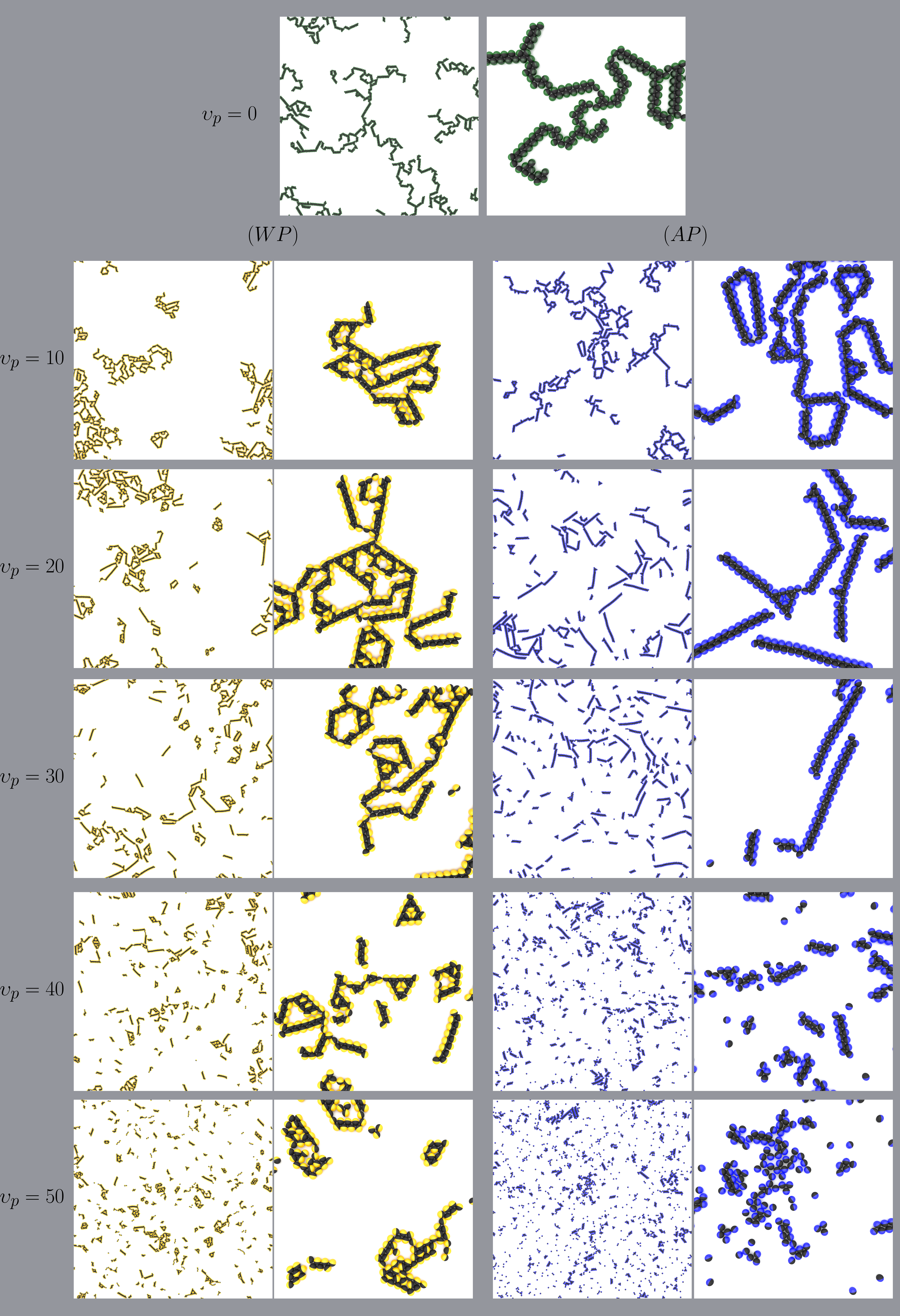}
  \caption{Typical simulation snapshot for active patchy particles with patch coverage $C=0.50$ and $\varepsilon=20$. The passive reference system $\upsilon_p=0$ is shown at the top of the figure.  The (WP) case is shown in the left column while the (AP) case is shown in the right column. Each system snapshot is accompanied by an enlarged snapshot highlighting the local cluster geometry.}
  \label{fig:6}
\end{figure}
\clearpage
}

The effects of self-propulsion are not as simple to unravel as in the $C=0.25$ case, where self-propulsion can either stabilizes or destabilizes a cluster depending on the direction of self-propulsion relative to the attractive patch. In Figure \ref{fig:4}(b), we show the average degree of aggregation $\langle \Theta \rangle$ as a function of activity for both binding energies considered.  In both cases,  $\langle \Theta \rangle$   decreases as $\upsilon_p$ is increased, and in both cases, the $(AP)$ direction experiences a significantly faster decay of $\langle \Theta \rangle$. Thus, one would naively concluded that the main effect of activity is that of reducing the average size of the aggregates in solution until a fluid state is recovered for very large $\upsilon_p$.  This line of thought also suggests that larger active forces are necessary in the $(WP)$ direction to break the clusters with respect to the $(AP)$ direction. However, this is not the entire story: a closer look at the evolution of the different configurations that develop in the two propelling directions reveals a complex structural behavior. Figures~\ref{fig:5} and~\ref{fig:6} present steady-state snapshots of typical structure that develop as activity $\upsilon_p$ is increased for $\varepsilon=12$ and $\varepsilon=20$, respectively.

A binding energy of $\varepsilon=12$ is somewhat ideal for the passive system, as the APPs self-assemble into long linear aggregates with only a sporadic formation of kinks along the chain (Fig.~\ref{fig:5}) . No branching of the chains is observed at these binding energies. When activity is introduced in the $(AP)$ direction (right column of Figure \ref{fig:5}), the length of the structures decreases with increasing $\upsilon_p$, and the formation of kinks becomes less probable.  The system eventually devolves into a hot fluid for even larger values of $\upsilon_p$. When activity is introduced in the $(WP)$ direction (left column of Figure \ref{fig:5}), patchy particles feel a stronger attraction when in the condensed phase. This leads to the more frequent formation of kinks along the chain, but simultaneously internal torques develop at the location of the kinks that act as pivots around which the chain can fold. A comparison of the structures for $\upsilon_p=25$ in the $(WP)$ and $(AP)$ directions is quite revealing (lowest panels in Fig.~\ref{fig:5}). In the $(WP)$ direction most of the particles are involved in the formation of small, rigid and compact clusters, while in the $(AP)$ direction most particles are in a fluid phase. This is consistent with the data shown in Fig.~\ref{fig:4}(b).

A much more dramatic illustration of this effect can be observed for $\varepsilon=20$. Here, large branched clusters develop in the passive case (Fig.~\ref{fig:6}, top panel). As activity is introduced either in the $(AP)$ (right column) or in the $(WP)$ direction (left column), we observe an overall collapse of the structures. The collapse or folding points in these structure are those locations where kinks or branching exist.  The kink portions of the structure act as a fulcrum, and the two linear segments surrounding this fulcrum generate a force imbalance leading to the compaction and folding of the structure.  However, while these structures, regardless of the direction of the active force, tend to become more compact for small values of $\upsilon_p$, at intermediates values of $\upsilon_p$, we observe diverging structural behaviors. 
In both $(WP)$ and $(AP)$ cases the large branched aggregates systematically break down into smaller chuncks. This is because kinks and branching points are the structural weak points of these assemblies and thus more susceptible to be unraveled by the active forces. However, when considering the $(WP)$ direction, we observe  small compact structures where multiple linear pieces merge together into activity-stabilized open-mesh clusters, whereas in the $(AP)$ direction, we observe a suspension of linear chains. 
This phenomenon is most evident as soon as the active forces become comparable in strength to the dispersion forces $\upsilon_p \simeq 20$.
\begin{figure}[h!]
 \centering
    \includegraphics[width=0.9\textwidth]{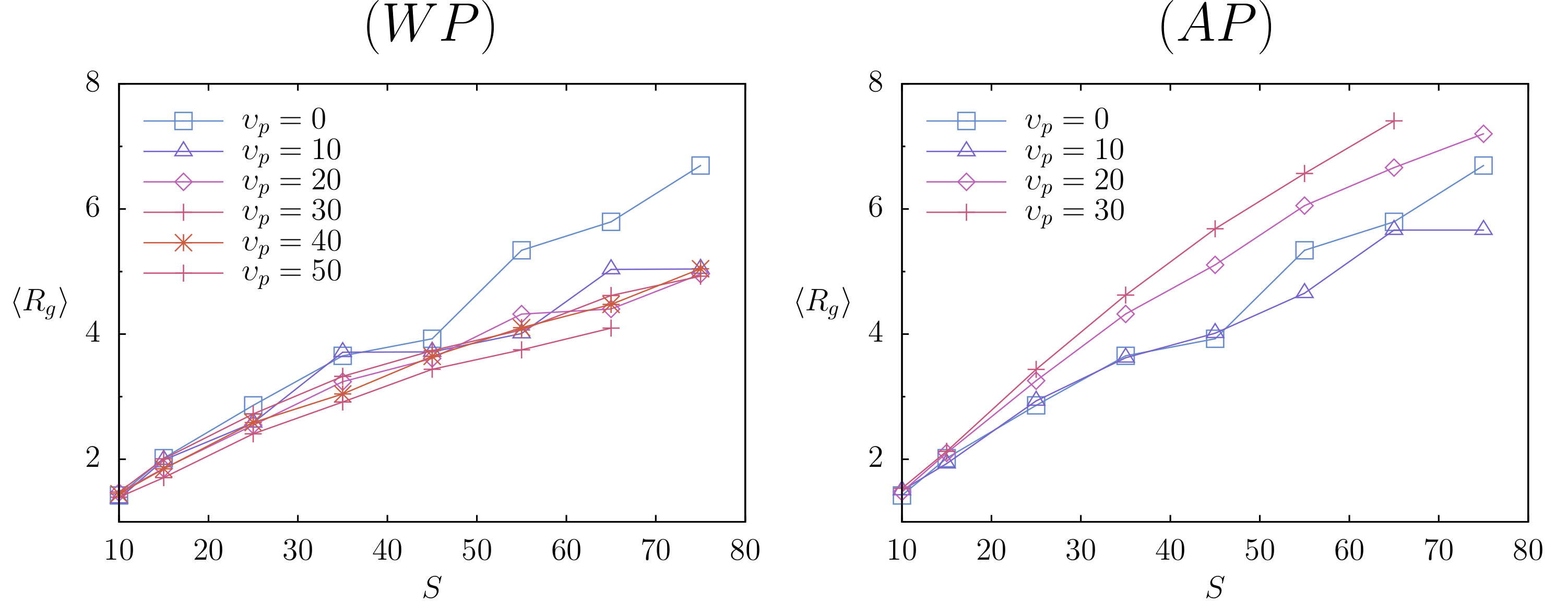}
  \caption{Average radius of gyration $\langle R_g \rangle$ as a function of cluster size $S$ for binding energy $\varepsilon=20$. The $WP$ case shows a  compaction of the clusters with increasing $\upsilon_p$, while the $AP$ case exhibits reentrant behavior where clusters undergo a collapse before an eventual expansion of $R_g$ as $\upsilon_p$ increases.}
  \label{fig:7}
\end{figure}

This diverging structural behavior for the two propelling directions can be more quantitatively expressed by measuring the average radius of gyration of the clusters formed by particle in the AP and WP directions. This analysis is shown in Figs.~\ref{fig:7}, where the average radius of gyration $\langle R_g \rangle$ is given as a function of clusters size $S$ for $\varepsilon=20$ at different propelling velocities $\upsilon_p$. Figure.~\ref{fig:7}(a) refers to the data for the $(WP)$ direction and illustrates how $R_g$ systematically decreases when going from the passive case, $\upsilon_p=0$ (blue curve), to the largest activity considered, $\upsilon_p=50$ (red lowest curve).  The $(AP)$ direction is given in figure.~\ref{fig:7}(b) and shown how $\langle R_g \rangle$ initially slightly decreases when compared to the passive reference system, but eventually moves up to larger values for sufficiently large activities (already for $\upsilon_p=20$). Unfortunately, it is very hard to obtain these data with sufficient statistics and for large enough clusters to extract a reliable size exponent. Nevertheless the two trends are quite clear. Structurally speaking, apart from the decreasing overall size of the clusters with the increase of $\upsilon_p$, one can think of the aggregates formed by the $(WP)$ particles as polymers in a bad solvent with a linear persistence length consisting of about 10 particles, and of the aggregates formed by the $(AP)$ particles, at least beyond the re-entrant point, as semiflexible polymers. 
Lastly, it is also worth noting that the activation of the APPs in the (AP) direction has a practical application as a method to wash away unwanted kinks and branches from a configuration of passive particles assembling with interactions that are too sticky.
 
\subsection{Dilute suspension of active APPs with patch coverage $(C=0.75)$}

\begin{figure}[h!]
 \centering
    \includegraphics[width=1.0\textwidth]{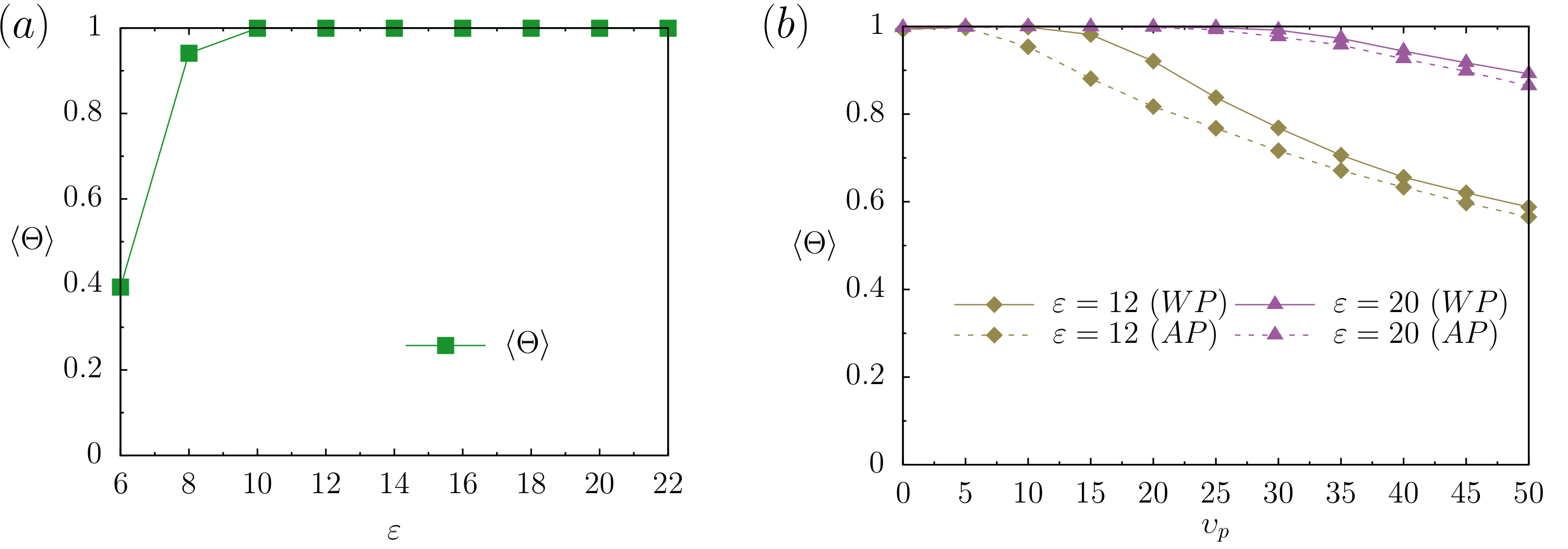}
  \caption{(a) Average degree of aggregation $\langle \Theta \rangle$  for passive suspensions of patchy particles with patch coverage $C=0.75$ as a function of the binding energy $\varepsilon$. (b) Average degree of aggregation $\langle \Theta \rangle$ as a function of $\upsilon_p$ for active suspensions with $C=0.75$. Two different binding energies are considered: $\varepsilon=12$ (light green) and $\varepsilon=20$ (magenta). Solid lines correspond to the (WP) case, while the (AP) case is indicated by the dotted lines.  
  }
  \label{fig:8}
\end{figure}

\begin{figure}[h!]
 \centering
    \includegraphics[width=0.75\textwidth]{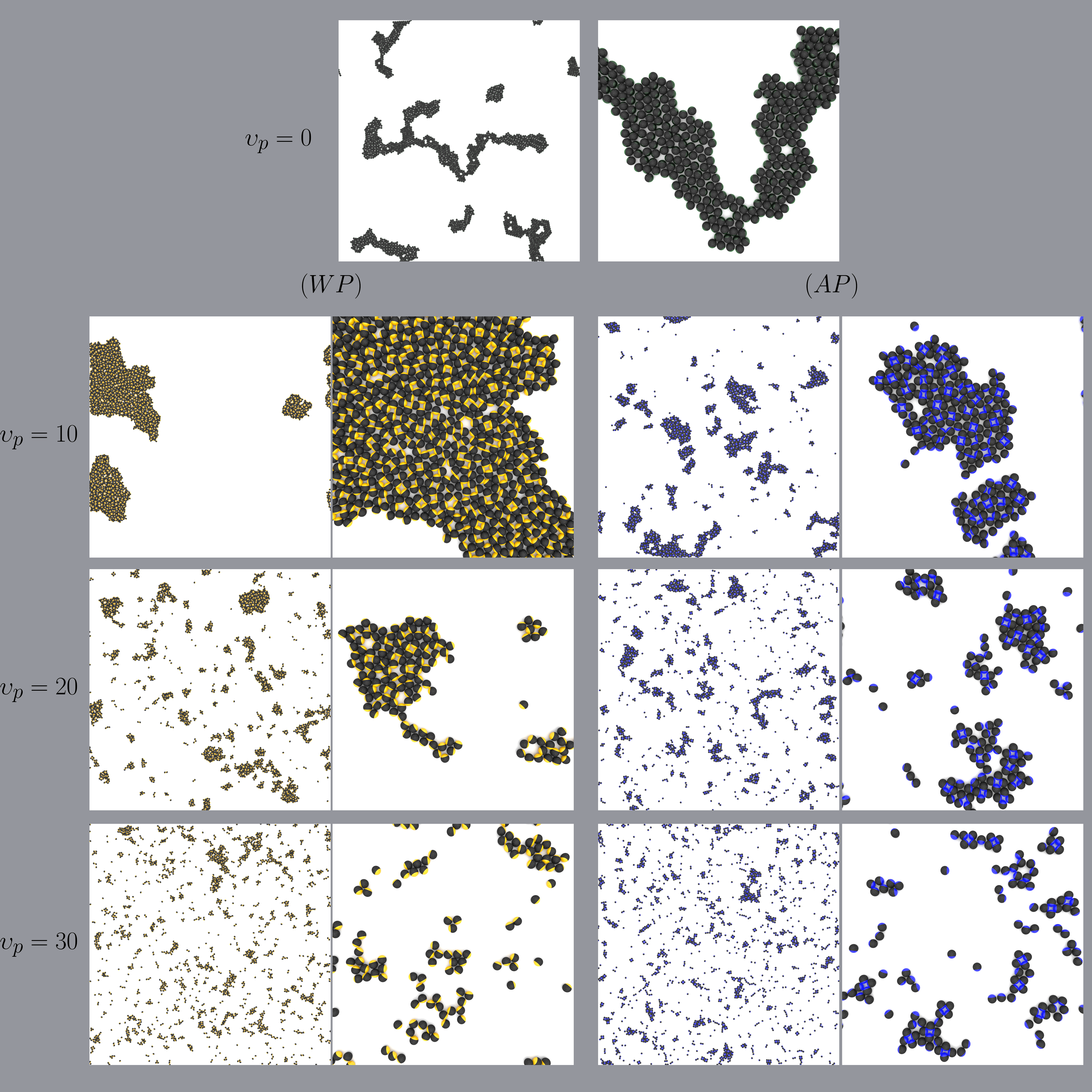}
  \caption{Typical simulation snapshot for active APPs with patch coverage $C=0.75$ and $\varepsilon=12$. The passive reference system $\upsilon_p=0$ is shown at the top of the figure.  The $(WP)$ case is reported in the left column whereas the $(AP)$ case is in the right column.  Each snapshot is accompanied by an enlarged snapshot highlighting the local cluster geometry.}
  \label{fig:9}
\end{figure}

We conclude this study by analyzing the bulk properties of a low density suspension of APPs with patch coverage $C=0.75$.  At this large patch coverage the geometry of the interaction is wide enough that each APP can accommodate more than 4 other APPs. This leads to the formation of more isotropic aggregates. In Fig.~\ref{fig:8}(a), we show $\langle \Theta \rangle$  as a function of the binding energy $\varepsilon$ for the passive system. The system fully aggregates into a single cluster for any $\varepsilon$ greater than about $10$, and for smaller values of $\varepsilon$, we observe small isotropic clusters with overall hexagonal symmetry. At $\varepsilon=12$, we report the formation thick crystalline branched structures.  See top panels of Fig.~\ref{fig:9} for a typical conformations of the passive suspension with $C=0.75$ and $\varepsilon=12$. Figure ~\ref{fig:8}(b)  shows the average  degree of aggregation $\langle \Theta \rangle$ as a function of activity for binding energies $\varepsilon=12$ and $\varepsilon=20$. Notably, when compared to the two smaller patch geometries considered in this study the difference in $\langle \Theta \rangle$ between the $AP$ and $WP$ directions is far less dramatic.  For both binding energies, $\langle \Theta \rangle$  decrease monotonically with increasing self-propelling velocity $\upsilon_p$. 

\begin{figure}[h!]
 \centering
    \includegraphics[width=0.75\textwidth]{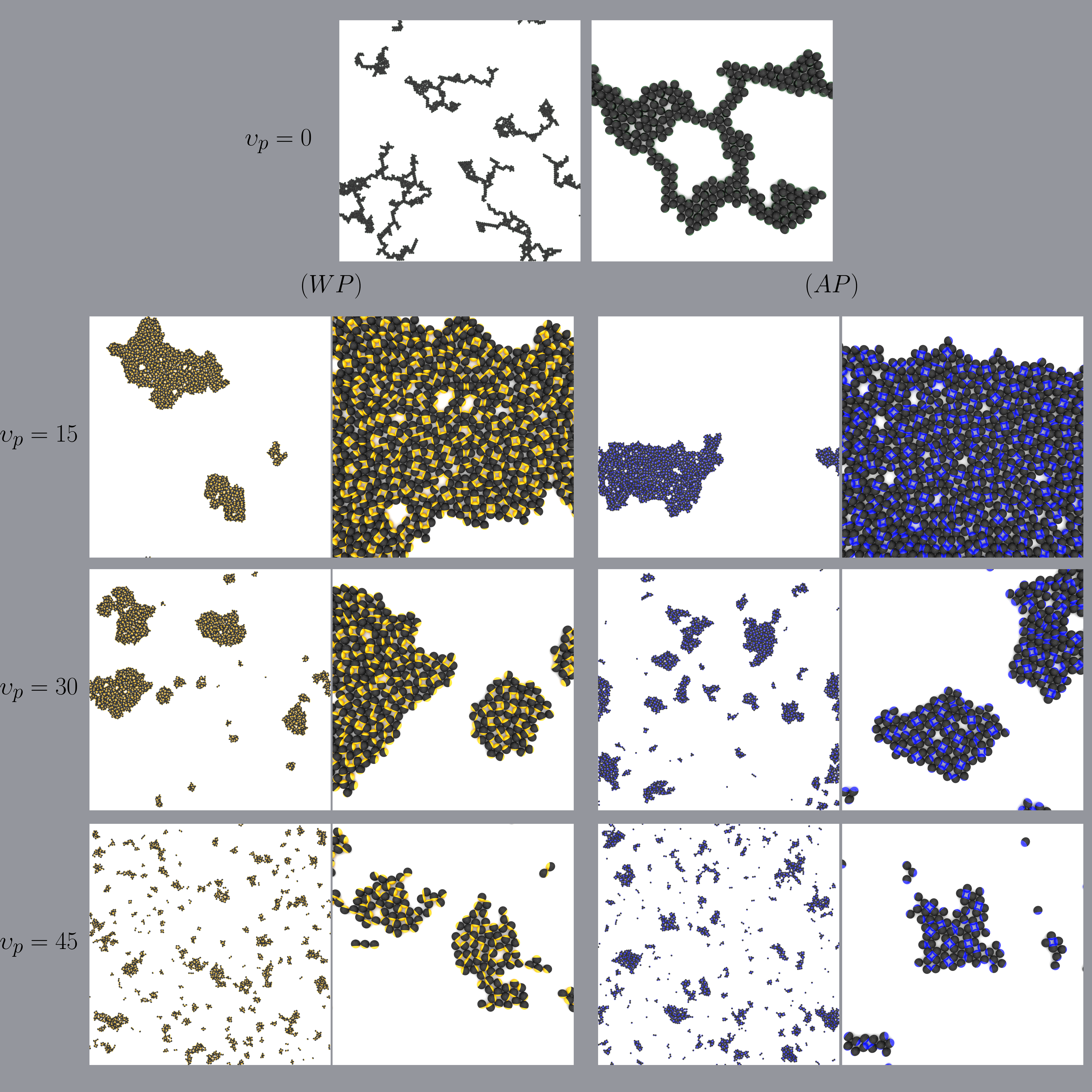}
  \caption{Typical simulation snapshot for active APPs with patch coverage $C=0.75$ and $\varepsilon=20$. The passive reference system $\upsilon_p=0$ is shown at the top of the figure.  The $(WP)$ case is shown in the left column while the $(AP)$ case is shown in the right column.  Each system snapshot is accompanied by an enlarged snapshot highlighting the local cluster geometry.}
  \label{fig:10}
\end{figure}

One of the most compelling features of the clusters formed both at $\varepsilon=12$ and at $\varepsilon=20$ is the change in cluster shape as soon as activity is introduced. Unlike the typical branched structures formed by the passive system, clusters formed  by active APPs immediately collapse into truly two dimensional objects with a size exponent equal to $1/2$,  which means that system form isotropic clusters, size exponent of $1/2$ comes from the calculation of the radius of gyration as a function of the cluster size. This is due in large part to the increased degree of rotational freedom of an APP allowing for more efficient folding of the cluster. Figures~\ref{fig:9} (lower panels) and ~\ref{fig:10} (lower panels) show the typical aggregates formed for binding energy $\varepsilon=12$ and $\varepsilon=20$, respectively. Apart from the slightly faster decay in $\langle \Theta \rangle$ for the $(AP)$ direction with respect to the $(WP)$ direction, there aren't significant structural differences across the range of active velocities considered in this study.

What is, however,  significant is the dynamical nature of the clusters themselves. In fact, the clusters that form are active in their own right, and they "move, break apart and form again" over the life-time of the simulations.  The behavior here is reminiscent of the behavior observed in experiments with attractive colloidal swimmers~\cite{palacci_living_2013}. The main difference between our system and that with active spherical particles interacting via an isotropic attractive potential is that in our case, the axis of propulsion of the particles in the condensed phase is essentially frozen, and thus patchy particles in the cluster do not rotate significantly. This is in stark contrast with the isotropic attractive case where active particles undergo rotational Brownian motion. For APPs, what drives these clusters unusual dynamics is their orientational and translational frustration. In fact, far from being hexagonal, these clusters are frozen into rather disordered configurations with frequent gaps within the structure and with the presence of several tetramers with square symmetry organized in such a way that the hydrophilic, rather than the hydrophobic sides of the particles point toward each other. The large torques and forces that develop within these clusters can therefore easily destabilize them as soon as they grow to a sufficiently large size, or, as we often observe, as soon as a small cluster in solution binds to it forcing orientational re-arrangements of the particles that can destabilize their already precarious balance of forces. Finally, it is worth mentioning that while living crystals observed for $C=1.00$ are typically in dynamic equilibrium with a fluid of unpaired particles, at low velocities, we observe fluids of active clusters, with no unpaired APPs, where clusters either break upon interacting with other clusters, or after large and slow internal cluster re-arrangements. 

\section{Discussion and conclusions}\label{Sec:conclusions}

\begin{figure}[h!]
 \centering
    \includegraphics[width=0.9\textwidth]{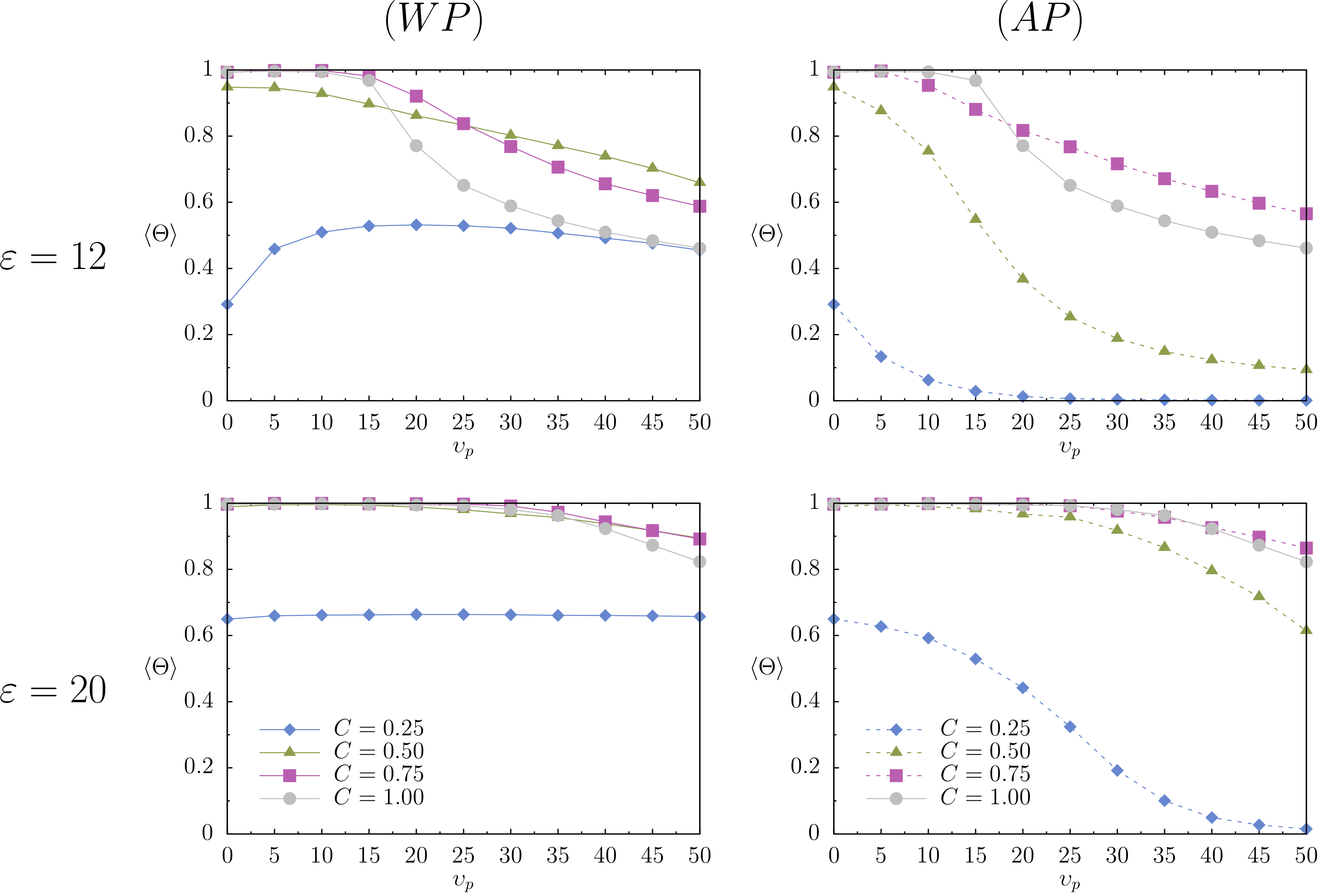}
  \caption{Summary of the Average degree of aggregation $\langle \Theta \rangle$ for all patch coverages considered in this study. The left hand side plots show the results for the $WP$ case, while the 
  plots on the right hand side show the results for the $AP$ case. As a reference, these plots also include the data for fully isotropic attractive particles corresponding to  $C=1$.}
  \label{fig:11}
\end{figure}

In this paper, we analyzed how active forces can be used to stabilize or destabilize aggregates formed by active amphiphilic patchy particles (APPs). We study the phase behavior of dilute suspensions of these patchy particles for three different hydrophobic patch coverages: $C=0.25$, $C=0.50$, and $C=0.75$. We also considered explicitly the role of the direction of the active forces with respect to the placement of the attractive patch, i.e. with the patch ($WP$) and against the patch ($AP$) for a range of different swimming velocities $\upsilon_p$ and hydrophobic attractions $\varepsilon$. Figure~\ref{fig:11} nicely summarizes all of our results concerning how the average degree of aggregation $\langle \Theta \rangle$  depends on the swimming velocity $\upsilon_p$ for two different values of $\varepsilon$. The left column shows results for the $(WP)$ direction while the right column corresponds to the $(AP)$ direction. For completeness we also added to these plots the data corresponding to the fully hydrophobic particles (isotropic attraction) for which $C=1.00$.

Our study illustrates how the geometry of the interactions between the particles plays a pivotal role in how active forces affect the stability of self-assembled aggregates. Unlike the very special case of spherical active particles interacting with an isotropic potential, where the rotational degrees of freedom of each particle aren't lost when in the condensed phase, using active APPs with a tunable size of the hydrophobic region enables us to systematically control the parameters responsible for their rotational dynamics. Interestingly, we find that under strong angular confinement $-$ condition that leads to the formation of small finite size clusters $-$ it is possible to increase the particle's aggregation yield by aligning the active forces in the direction of the patch $(WP)$. We have also shown how, when only half of the particle is covered by a hydrophobic patch, it is possible to break apart and either compactify or wash away branches from large branched networks of particles. In doing so we highlighted the significant role played by the direction of the propelling forces for particle with 50\% hydrophobic coverage. Finally, we discussed the remarkable behavior of the isotropic active clusters that self-assemble when 75\% of the particle area is covered by a hydrophobic patch, and point out how frustration of the frozen active forces within the clusters leads to their dynamic, "living" behavior. One can think of these aggregates as being in between those formed by active attractive rods, and those formed by isotropic active Lennard Jones particles. In fact, in clusters formed by rods (or dumbbells), particles are unable to rotate even when a second cluster merges with it. This leads to the formation of ever growing macroscopic aggregates that have a net rotational speed that depends on the details of the particle's relative orientations, and that decreases with the cluster size~\cite{schwarz-linek_phase_2012}. In clusters formed by spherical particles, as mentioned above, the colloids are completely free to rotate within the clusters, and for a range of active forces and for sufficiently low densities, this leads to the formation of "living" clusters which do not seem to grow beyond a certain size~\cite{palacci_living_2013}. Active APPs with 75\% coverage form clusters where the directions of the active forces of the particles is mostly frozen, leading to rotating clusters with a net rotational speed, but in this case the propelling forces can undergo significant re-arrangements when interacting with another cluster, leading to cluster fractures, large global re-arrangements and eventual disassembly.

Hydrodynamic interactions may play an important role on the phase behavior and structural properties of the assemblies formed by these particles. Extensive work on the subject is currently underway and it will be published elsewhere.

\section*{Acknowledgements} 
We thank Clarion Tung and Joseph Harder for insightful discussions and helpful comments. C.V. acknowledges financial support by a Ramon y Cajal tenure track, the Spanish Ministry of Education (MINECO, FIS2016-78847-P) and UCM-Santander (PR26/16-10B-3). A.C. acknowledges financial supported from the National Science Foundation under Grant No. DMR-1408259. SAM acknowledges financial support from the National Science Foundation Graduate Research Fellowship. This work used the Extreme Science and Engineering Discovery Environment (XSEDE), which is supported by National Science Foundation grant number ACI-1053575.

%%%\bibliography{references}

%

\end{document}